\documentstyle[aps,amstex,amssymb,preprint,epsf]{revtex}
\tightenlines
\begin{document}                                                       

\draft 
                                                   
\title{MAPPING OF EUCLIDEAN RESONANCE\\ 
ON STATIC RESONANT TUNNELING}

\author{B. Ivlev} 
         
\address
{Department of Physics and Astronomy\\
University of South Carolina, Columbia, SC 29208\\
and\\
Instituto de F\'{\i}sica, Universidad Aut\'onoma de San Luis Potos\'{\i}\\
San Luis Potos\'{\i}, S. L. P. 78000 Mexico}

%\date{\today}
\maketitle

\begin{abstract}
Quantum tunneling through an almost classical potential barrier can be strongly enhanced by a nonstationary field so that the penetration
through the barrier becomes not exponentially small. This constitutes an extremely unusual phenomenon of quantum physics called Euclidean 
resonance. A certain nonstationary barrier is proposed with a very low WKB tunneling rate. The quantum dynamics of this barrier is 
mapped on resonant tunneling across a static double barrier with a resonant level inside. The real penetration through the dynamical 
barrier is not exponentially small providing an example of Euclidean resonance. Therefore, the Schr\"{o}dinger equation allows solutions of
the type of Euclidean resonance. The counterintuitive phenomenon of Euclidean resonance is a dynamical analogue of static resonant 
tunneling. 

\end{abstract} \vskip 1.0cm
   
 
\narrowtext

\section{INTRODUCTION}
\label{sec:intro}
The problem of quantum tunneling through nonstationary potential barriers was addressed in Refs.\cite{KELDYSH} and \cite{PERELOMOV1}. The 
method of complex classical trajectories was developed in Refs.\cite{MELN1,MELN2,MELN3,MELN4}. Recent achievements in semiclassical theory 
are presented in Refs. \cite{KESHA,BERMAN,DEFENDI,MAITRA,ANKERHOLD1,CUNIBERTI,ANKERHOLD2,ANKERHOLD3,POLLAK}. See also the related papers 
\cite{DYAK,ZEL,GURVITZ,HANGGI}. In Refs.\cite{IVLEV1,IVLEV2} the advanced approach was developed to go beyond the method of classical 
trajectories and to obtain a space-time dependence of the wave function in semiclassical regime. 

Tunneling of a particle through a static barrier is described by the semiclassical theory of Wentzel, Kramers, and Brillouin (WKB) 
\cite{LANDAU}. A motion through a nonstationary barrier is commonly treated as a combination of quanta absorption and tunneling. The 
particle pays in probability to absorb quanta and to reach the certain higher energy level but the subsequent tunneling is easier since it 
occurs in a more transparent part of the barrier. This mechanism of barrier penetration in a nonstationary field has no conflict with 
intuition and is called photon-assisted tunneling. 

Besides photon-assisted tunneling, a penetration through a nonstationary barrier can occur according to a scenario of a completely 
different physical nature. Briefly, the basis of this phenomenon is a formation of under-barrier dynamical state which starts to move away 
of the barrier and carries a non-exponentially small fraction of the incident wave packet. This phenomenon is called Euclidean resonance 
(ER) since an outgoing flux depends sensitively on dynamical parameters \cite{IVLEV2,IVLEV3,IVLEV4}. The same refers to an amplitude of the
under-barrier state which can be called a resonant state. 

The phenomenon of Euclidean resonance can be manifested in any physical process where tunneling is a substantial part. In particular, one 
can consider ER application for selective disintegration of atoms and molecules through artificially created energy barriers by applying 
d.c. and a.c. fields \cite{IVLEV4}. The potential areas for ER applications are scanning tunneling microscopy (STM) 
\cite{STROSCIO,GOMER,GRAF}, nuclear physics \cite{WEISSKOPF,IVLEV3}, molecular electronics \cite{SEM}, nanoscience, tunneling chemical 
reactions \cite{MIYA}, and quantum mechanics of Josephson junctions \cite{BARONE,CLARKE,USTINOV}. 

In contrast to photon-assisted tunneling, the phenomenon of Euclidean resonance is counterintuitive. This means that it is very hard to 
reduce it to a combination of known physical effects. Euclidean resonance was established on the basis of solution of the Schr\"{o}dinger
equation by methods of classical trajectories in complex time \cite{IVLEV2,IVLEV3,IVLEV4} and by exact solution of Hamilton-Jacobi
equation for the classical action (semiclassical approximation \cite{IVLEV2}). The last approach was mostly advanced but it was also not an 
exact mathematical solution of the whole problem since semiclassical approximation violated at the moment of formation of the under-barrier
state and one had to go around that singular point. In this situation, any interpretation of Euclidean resonance through a set of known 
physical effects would be very useful. Unfortunately, such interpretation in the case of a general nonstationary barrier was not easy. 

Another way, to interpret Euclidean resonance, is to point out {\it at least one} particular nonstationary potential for which one can 
obtain a solution of the Schr\"{o}dinger equation of ER type. Then, on the basis of this mathematical solution, a physical interpretation 
would be possible. The nonstationary potential may be very specific and not necessary having a direct physical application. It is not 
important in this case since one should answer the general and non-trivial question: Does the Schr\"{o}dinger equation allow a solution of 
ER type in principal? The answer is ``yes'' if one can find at lest one example of this type. The nonstationary potential to be found 
should be at all times non-transparent in the WKB sense and slow varying to avoid an over-barrier excitation of the particle.  

The above program has been accomplished in the paper. With the nonstationary potential proposed the dynamical problem is mapped on well
known resonant tunneling through a static double barrier system with a resonant level inside. When the energy of an incident wave coincides
with the resonant level the transition probability becomes non-exponentially small. Using this example, one can follow a formation of the 
resonant state inside the barrier system. One can also observe that the formation of the resonant state is a rigorously non-semiclassical 
process, related to a strong quantum coherence, as was stated before \cite{IVLEV2,IVLEV4}. So, the dynamical potential, considered in the 
paper, reproduces the main features of Euclidean resonance. 

One can conclude that Euclidean resonance is a dynamical analogue of static resonant tunneling.

In Sec.~\ref{general} general mechanisms of barrier penetration are mentioned. In Sec.~\ref{static} resonant tunneling through static double
barrier system is reviewed. In Secs.~\ref{dynamics} - \ref{acceler} details of the dynamical potential are given. 
\section{HOW A PARTICLE PENETRATES THROUGH A DYNAMICAL BARRIER}
\label{general}
In this Section we briefly analyze different types of quantum tunneling through potential barriers.
\subsection{WKB tunneling} 
In classical physics a particle cannot penetrate through a potential barrier. In contrast to this, quantum mechanics allows a finite 
probability of such transition. According to WKB theory, the probability $\exp\left[-A(E)\right]$ to penetrate through a static barrier 
depends on a particle energy $E$ \cite{LANDAU}. For almost classical potential barriers the value $A(E)$ is big and WKB tunneling 
probability is exponentially small. 
\subsection{Photon-assisted tunneling}
When a potential barrier is acted by a nonstationary field another phenomenon enters the game. A particle can go through a barrier by a 
combined mechanism. Schematically, this mechanism is a simultaneous absorption of quanta of nonstationary field and subsequent tunneling. 
This process is called photon-assisted tunneling.

Suppose a particle to have the initial energy $E$. Then it can absorb $N$ quanta of the energy $\hbar\Omega$ each and to tunnel in a more 
transparent part of the barrier with the higher energy $E+N\hbar\Omega$. The total probability of this combined process 
$w^{2N}\exp\left[-A(E+N\hbar\Omega)\right]$ should be extremized with respect to the number $N$ of absorbed quanta. Here $w$ is a 
probability to absorb one quantum which is proportional to a nonstationary field. 

The resulting probability depends on the value of $w$. When $w$ is small, it may happen that the optimal $N=0$ and then the probability is
mainly $\exp\left[-A(E)\right]$ which it would be in the absence of the nonstationary field. In this case, one or two quanta absorption has
lower probability providing only a weak assistance of tunneling. When $w$ is not very small the number $N$ of absorbed quanta may be big 
and this multiquanta assistance of tunneling has essentially bigger probability compared to the static result $\exp\left[-A(E)\right]$. 

The probability of photon-assisted tunneling is not exponentially small only if a particle can effectively reach the barrier top (no 
tunneling component in the process). This can happen, for example, when the nonstationary field is not small or its typical frequency 
$\Omega$ is of the order of a barrier height.  

In principal, the idea of photon-assisted tunneling is clear and this mechanism was commonly considered as a unique one in physics of 
quantum mechanical motion through potential barriers. 
\subsection{ Euclidean resonance}
\begin{figure}[p]
\begin{center}
\vspace{1.5cm}
\leavevmode
\epsfxsize=\hsize
\epsfxsize=16cm
\epsfbox{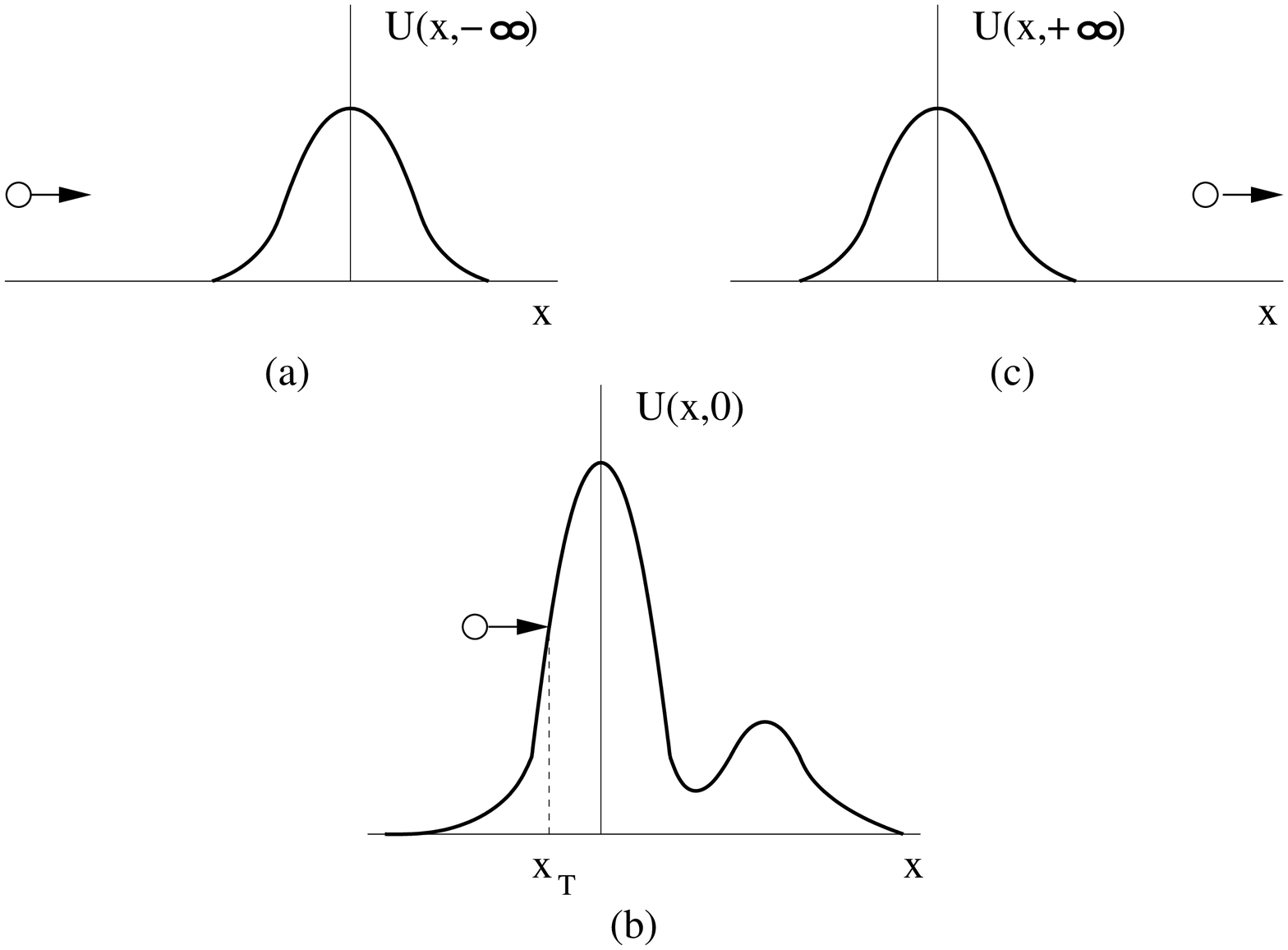}
\vspace{1cm}
\caption{A particle (a wave packet) moves through the slow varying potential $U(x,t)$. At the moment $t=0$ the particle collides the 
barrier at the classical turning point $x_{T}$.}
\vspace{4cm}
\label{fig1}
\end{center}
\end{figure}

Suppose, a particle moves in the nonstationary potential $U(x,t)$ which has, generally, a shape of a barrier as shown in Fig.~\ref{fig1}.
This potential can be characterized by its intrinsic frequency $\omega$ which is simply of the order of a frequency of classical oscillation
in the over-turned barrier $-U$. The nonstationary potential $U(x,t)$ is supposed to vary adiabatically. This means that its typical 
frequency is much smaller than $\omega$. An adiabatic barrier is considered in this paper only for convenience. In analogous problems a 
typical frequency can be of the order of $\omega$ or even bigger. 

The barrier $U(x,t)$ is almost classic at all times in the sense that WKB tunneling probability $\exp(-A)$ is always small 
($A\sim U/\hbar\omega$ is big). The incident wave packet (particle) in Fig.~\ref{fig1}(a) can be described by the wave function 
\cite{FEYNMAN}
\begin{equation}
\label{1}
\psi(x,t)\sim\exp\left[\frac{i}{\hbar}S(x,t)\right]
\end{equation}
where $S(x,t)$ is a classical action. The wave function (\ref{1}) reaches a maximal value at the classical trajectory $x(t)$ corresponding 
to the extremal action $S\left[x(t),t\right]$. The classical trajectory obeys Newton's equation
\begin{equation}
\label{2}
m\frac{\partial^{2}x}{\partial t^{2}}+\frac{\partial U(x,t)}{\partial x}=0
\end{equation}
The wave packet moves as a classical particle $x(t)$ in the adiabatically varying potential $U(x,t)$ and collides the barrier at the moment
$t=0$ at the classical turning point $x=x_{T}$ as indicated in Fig.~\ref{fig1}(b). At the turning point the semiclassical approximation
(\ref{1}) for the incident wave function breaks down. 

What happens further?  

The potential $U(x,t)$ varies slow compared to the intrinsic time $1/\omega$ of the system. Therefore, a natural assumption for the further
particle motion is a conventional WKB tunneling with the energy $U\left(x_{T},0\right)$ through the static potential barrier $U(x,0)$ in  
Fig.~\ref{fig1}(b). This obvious mechanism for sure exists but it is not a unique one. 

A penetration through a nonstationary barrier can also occur according to a scenario of a completely different physical nature. Briefly, 
the basis of this phenomenon is a formation of the under-barrier dynamical state which starts to move away of the barrier, as in 
Fig.~\ref{fig1}(c), and carries a non-exponentially small fraction of the incident wave packet. Despite $U(x,t)$ varies slow, the formation
of the under-barrier state occurs much faster than the intrinsic time $1/\omega$. This phenomenon is called Euclidean resonance since an 
outgoing flux depends sensitively on dynamical parameters \cite{IVLEV2,IVLEV3,IVLEV4}. The same refers to an amplitude of the under-barrier
state which can be called a resonant state. 

The phenomenon of photon-assisted tunneling allows to be treated as a sequence of partial processes of photon absorption and tunneling. In
contrast to this, Euclidean resonance was impossible to be explained as a combination of some known effects. Now in this paper a particular
dynamical barrier is proposed which allows to make a link between Euclidean resonance and some phenomena which are already known in quantum
physics. 

The problem to be solved is to construct a barrier which satisfies the conditions
\begin{align}
\label{3a}
&{\rm \hspace{0.15cm}very\hspace{0.15cm}slow\hspace{0.15cm}variation\hspace{0.15cm}in\hspace{0.15cm}time}\\
\label{3b}
&{\rm \hspace{0.15cm}exponentially\hspace{0.15cm}small\hspace{0.15cm}WKB\hspace{0.15cm}penetration\hspace{0.15cm}at\hspace{0.15cm}all 
\hspace{0.15cm}times}\\
\label{3c}
&{\rm \hspace{0.15cm}real\hspace{0.15cm} penetration\hspace{0.15cm} through\hspace{0.15cm}the\hspace{0.15cm}barrier\hspace{0.15cm}is 
\hspace{0.15cm}{\it not}\hspace{0.15cm}exponentially\hspace{0.15cm} small}
\end{align}
\section{STATIC RESONANT TUNNELING}
\label{static}
\begin{figure}[p]
\begin{center}
\vspace{1.5cm}
\leavevmode
\epsfxsize=\hsize
\epsfxsize=16cm
\epsfbox{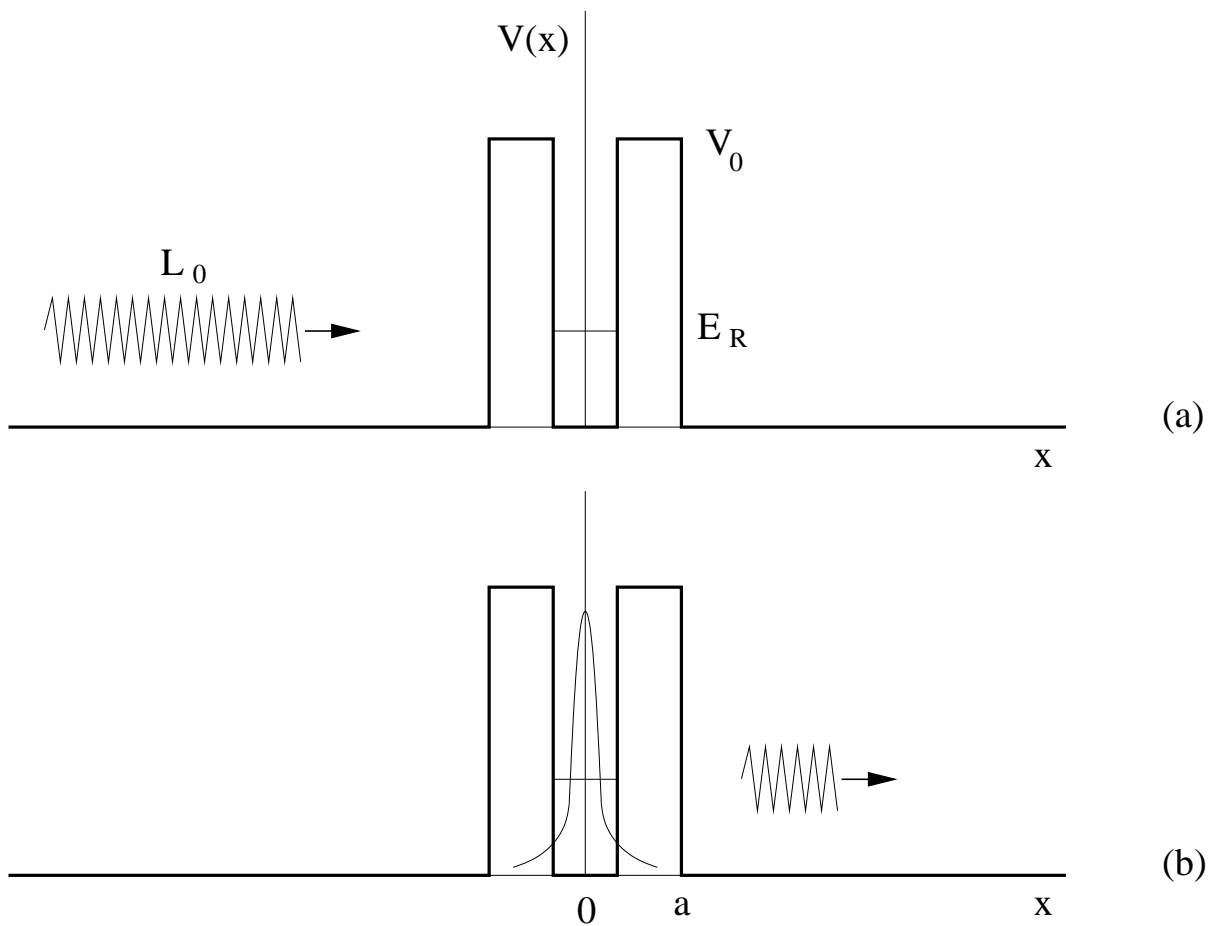}
\vspace{1cm}
\caption{Penetration of a wave packet through the symmetric double barrier static system as an illustration of static resonance tunneling. 
(a) the energy of the incident wave coincides with the resonant level $E_{R}$ in the well and the length $L_{0}$ of the incident packet is 
long. (b) for these conditions a non-exponentially small fraction of the incident flux penetrates the barrier. The resonant level has a big
occupation shown by the thin curve.}
\label{fig2}
\end{center}
\end{figure}

Let us consider first the symmetric double-barrier potential $V(x)$ shown in Fig.~\ref{fig2}. Each barrier is almost classical with the 
exponentially small tunneling probability $\exp(-A)$ which is determined by the WKB formula \cite{LANDAU}. In other words, $\exp(-A)$ is a 
tunneling probability through the barrier which is $V(x)$ at $x<0$ and equals $V(0)$ at $0<x$. Since the both barriers are weakly 
transparent one can consider quasi-discrete (with a small width proportional to $\exp(-A)$) energy levels in the well between them. The 
level $E_{R}$ is shown in Fig.~\ref{fig2}. 

If there is a steady incident flux of particles, with the energy $E$, from the left to the barriers $V(x)$ then the penetrated flux at 
$x\rightarrow +\infty$ strongly depends on $E$. When $E$ does not coincide with any discrete level tunneling processes through two barriers
are not coherent. In this case the total tunneling probability through the double barrier $V(x)$ is proportional to the product of partial 
probabilities $\exp(-A)\exp(-A)=\exp(-2A)$. 

Under the condition $E=E_{R}$, one can say that a particle multiply reflects in the well between the barriers before to tunnel through the 
right one in Fig.~\ref{fig2}(b). This results in strong coherence between tunneling processes through two barriers and the total tunneling 
probability through the potential $V(x)$ equals unity. This is called resonant tunneling since the process goes through the resonant state 
$E_{R}$. The idea of resonant tunneling steams to Wigner \cite{LANDAU}. One should emphasize that this scenario holds in a static situation
when incident and penetrated fluxes are steady. In this case the resonant level $E_{R}$ is highly occupied with the density 
$|\psi_{R}|^{2}\sim\exp(A)|\psi(-\infty)|^{2}$. 

When an incident flux is not steady but is a wave packet of a finite length $L_{0}$, as in Fig.~\ref{fig2}(a), the situation with quantum
coherence is different. The energy uncertainty of the incident wave packet $\hbar^{2}k\Delta k/m$ should not exceed the width 
$E_{R}\exp(-A)$ of the energy level $E_{R}$. Since $\hbar^{2}k^{2}/2m=E_{R}$ the minimal length $L_{0}\sim 1/\Delta k$ and the minimal 
duration $T_{0}\sim L_{0}/v$ of a wave packet obey the relations ($mv=\hbar k$)
\begin{equation}
\label{5}
L_{0}\sim\frac{\hbar}{\sqrt{mE_{R}}}\exp(A)\hspace{2cm}T_{0}\sim\frac{\hbar}{E_{R}}\exp(A)
\end{equation}
The duration $T_{0}$ of the wave packet is long which is sufficient to pump the resonant level up to the high occupation $|\psi_{R}|^{2}$. 
Then this high particle density on the level $E_{R}$ tunnels through the right-hand side barrier in Fig.~\ref{fig2}(b) providing the 
outgoing flux of the order $\exp(-A)|\psi_{R}|^{2}\sim|\psi(-\infty)|^{2}$. 

This means that the long wave packet, of the minimal length $L_{0}$ and of the wave number adapted to the discrete level 
$\hbar^{2}k^{2}/2m=E_{R}$, tunnels with the probability which is not exponentially small.

For a short wave packet, with a duration much less than $T_{0}$, the big uncertainty of the wave number in the wave packet $\Delta k$ leads
to the energy uncertainty $\hbar^{2}k\Delta k/m$ which exceeds the width of the resonant level $E_{R}\exp(-A)$. For this reason, a short 
wave packet tunnels through two barriers in a non-coherent way with the probability  of the order $\exp(-A)\exp(-A)=\exp(-2A)$. 

Suppose a wave packet moves towards the modified potential barrier $V(x/r)/r^{2}$ where $r$ is some small constant. In this case the WKB 
tunneling probability  $\exp(-A)$ through one barrier depends on almost the same $A$. The discrete level becomes now $E_{R}/r^{2}$ 
resulting, according to Eq.~(\ref{5}), in the packet length and duration
\begin{equation}
\label{6}
L\sim r\frac{\hbar}{\sqrt{mE_{R}}}\exp(A)\hspace{2cm}T\sim r^{2}\frac{\hbar}{E_{R}}\exp(A)
\end{equation}
If to chose the parameter $r\sim\exp(-A)$ a packet of a non-exponentially big length can penetrate through the barrier with 
non-exponentially small probability. Such a wave packet has the exponentially big energy $E_{R}/r^{2}\sim E_{R}\exp(2A)$. Its velocity, 
proportional to $\sqrt{E_{R}/m}\hspace{0.1cm}\exp(A)$, enables to pump up the resonant state within the short time $(\hbar/E_{R})\exp(-A)$.
One should note that static resonant tunneling cannot be described within semiclassical approximation since it substantially occurs through
a discrete level. 
\section{DYNAMICS OF THE BARRIER AND OF THE WAVE PACKET}
\label{dynamics}
As we have seen in Sec.~\ref{static}, a wave packet of a non-exponentially big length but with the exponentially high velocity 
$\sim\exp(A)$ behaves unusually. It can penetrate an exponentially shrunk double-barrier, of the width $\sim\exp(-A)$, and with the 
enhanced magnitude $\sim\exp(2A)$ during a time interval which is not exponentially long. 

Below we find a nonstationary potentials which accelerate a particle up to above high speed and which also shrink and magnify a 
double-barrier potential. In this case one can expect a barrier transition of a non-exponentially small probability.  

Let us consider the nonstationary potential of the form 
\begin{equation}
\label{7}
U(x,t)=\frac{1}{r^{2}(t)}V\left[\frac{x}{r(t)}\right]-x{\cal E}(t)
\end{equation}
with the double barrier potential $V$ as in Fig.~\ref{2} where the barrier amplitude is $V_{0}$. The function $r(t)$ provides a 
magnification and a shrinking of the potential barrier and the electric field ${\cal E}(t)$ is responsible for particle acceleration. The 
functions $r(t)$ and ${\cal E}(t)$ are plotted schematically in Fig.~\ref{fig3}. Shrinking and magnification occur close to the moments 
$\pm t_{0}$ where the electric field ${\cal E}(t)$ is almost zero. On the other hand, the electric field ${\cal E}(t)$ acts on background 
of the steady barrier $V(x/r_{0})/r^{2}_{0}$, where $r_{0}$ is exponentially small. In Secs.~\ref{shrink} and \ref{acceler} the both 
dynamical processes are discussed in details.
\section{SHRINKING AND MAGNIFICATION OF THE BARRIER}
\label{shrink}
According to Fig.~\ref{fig3}, in the time intervals $-t_{0}<t<-t_{1}$ and $t_{1}<t<t_{0}$ the electric field ${\cal E}(t)$ equals zero and 
the Schr\"{o}dinger equation with the potential (\ref{7}) has the form
\begin{equation}
\label{8}
i\hbar\hspace{0.1cm}\frac{\partial\psi(x,t)}{\partial t}=-\frac{\hbar^{2}}{2m}\hspace{0.1cm}\frac{\partial^{2}\psi}{\partial x^{2}}
+\frac{1}{r^{2}(t)}V\left[\frac{x}{r(t)}\right]\psi
\end{equation}
When a wave packet is localized to the left from the barrier, as in Fig.~\ref{fig2}(a), it moves mainly as a free particle and an influence
of the dynamical potential $V[x/r(t)]/r(t)^{2}$ is minor. Let us check this more carefully. 

The function $r(t)$ is even. In the vicinity of the point ($-t_{0}$) one can choose the particular function 
\begin{equation}
\label{9}
r(t)=\frac{1}{2}\left[1-\tanh\Omega(t+t_{0})\right]+\frac{r_{0}}{2}\left[1+\tanh\Omega(t+t_{0})\right]
\end{equation}
where the frequency $\Omega$ is smaller than the characteristic frequency 
\begin{equation}
\label{10}
\omega =\frac{\hbar^{2}}{ma^{2}}
\end{equation}
of the potential in Fig~\ref{fig2}. The moment of time $t_{0}$ can be chosen from the condition $\exp(-2\Omega t_{0})<r_{0}$. We apply the 
transformation of the wave function
\begin{equation}
\label{11}
\psi(x,t)=\frac{1}{\sqrt{r(t)}}\exp\left(\frac{im\dot{r}}{2\hbar r}x^{2}\right)
\Phi\left[\frac{x}{r(t)}\hspace{0.1cm},\hspace{0.1cm}T(t)\right]
\end{equation}
The transformation of the type (\ref{10}) was used in Ref.~\cite{PERELOMOV2} to study harmonic oscillator. The new time variable is defined
by
\begin{equation}
\label{12}
T(t)=\int^{t}_{-t_{0}}\frac{dt_{1}}{r^{2}(t_{1})}
\end{equation}
Then the wave function $\Phi(z,T)$ satisfies the equation
\begin{equation}
\label{13}
i\hbar\hspace{0.1cm}\frac{\partial\Phi(z,T)}{\partial T}=-\frac{\hbar^{2}}{2m}\hspace{0.1cm}\frac{\partial^{2}\Phi}{\partial z^{2}}+
\frac{m\Omega^{2}}{2}f(T)z^{2}\Phi +V(z)\Phi
\end{equation}
In Eq.~(\ref{13}) the function
\begin{equation}
\label{14}
f(T)=\frac{1}{\Omega^{2}}\hspace{0.1cm}\ddot{r}\left[t(T)\right]r^{3}\left[t(T)\right] 
\end{equation}
is of the order of unity and is localized on the time $T\sim 1/\Omega$. This can be easily seen from Eqs.~(\ref{10}) and (\ref{12}). 

Eq.~(\ref{13}) allows to follow quantum dynamics in a convenient way since the dynamical perturbation $v(z,T)=m\Omega^{2}f(T)z^{2}/2$ acts 
on the background of the static potential $V(z)$. If to put $V(z)=0$ than the dynamical perturbation corresponds to harmonic oscillator and
in coordinates $\{x,t\}$ there is simply a free motion. An inclusion of $V(z)$ results in quanta absorption in the spatial region where
$V(z)$ is localized. In the system $\{z,T\}$ the static potential $V(z)$ is localized far ahead of the moving wave packet during the action
of the dynamical perturbation $v(z,T)$ . Since $\hbar\Omega$ is much less than the barrier hight $V_{0}$ it takes the big number 
$N=V_{0}/\hbar\Omega$ quanta to reach the barrier top in order to provide photon-assisted transition. The probability of such transition 
\cite{MELN4}
\begin{equation}
\label{14a}
w^{2N}=\exp\left(-\frac{2V_{0}}{\hbar\Omega}\hspace{0.1cm}\ln\frac{1}{w}\right)\hspace{2cm}
\left(w=\frac{\lambda{\cal E}_{\rm ef}}{\hbar\Omega}\right)
\end{equation}
depends on the de Broglie length $\lambda =\hbar/\sqrt{mV_{0}}$ and the effective electric field 
${\cal E}_{\rm ef}\sim -\partial v/\partial z$ at the coordinate of the barrier jump in Fig.~\ref{fig2}. The parameter $w$ can be easily 
estimated as $w\sim (\Omega/\omega)\sqrt{\hbar\omega/V_{0}}\ll 1$. Therefore, the probability (\ref{14a}) of over barrier photon assistance 
is exponentially small. 

So, shrinking and magnification of the potential barrier at $t=-t_{0}$ weakly influences a process of barrier penetration resulting in an 
exponentially weak over-barrier photon assistance.  
\section{ACCELERATION OF THE WAVE PACKET}
\label{acceler}
\begin{figure}[p]
\begin{center}
\vspace{1.5cm}
\leavevmode
\epsfxsize=\hsize
\epsfxsize=16cm
\epsfbox{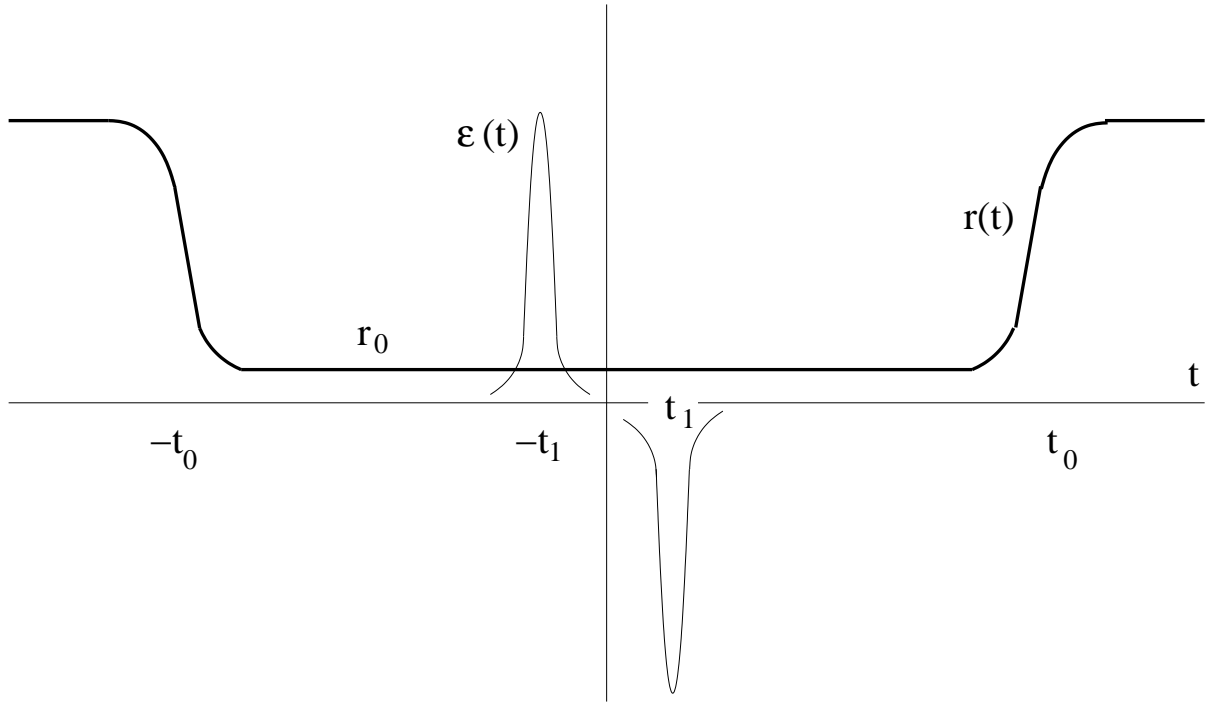}
\vspace{1cm}
\caption{The parameter $r(t)$ (solid curve) is responsible for shrinking and magnification of the barrier system at the time interval 
$\{-t_{0}, t_{0}\}$. The electric field ${\cal E}(t)$ (thin curve) provides a high speed of a particle at the interval 
$\{-t_{1}, t_{1}\}$.} 
\label{fig3}
\end{center}
\end{figure}

As follows from Fig.~\ref{fig3}, in the time interval $-t_{0}<t<t_{0}$ the quantum motion is described by the Schr\"{o}dinger equation
\begin{equation}
\label{15}
i\hbar\hspace{0.1cm}\frac{\partial\psi(x,t)}{\partial t}=-\frac{\hbar^{2}}{2m}\hspace{0.1cm}\frac{\partial^{2}\psi}{\partial x^{2}}
-x{\cal E}(t)\psi +\frac{1}{r^{2}_{0}}V\left(\frac{x}{r_{0}}\right)\psi
\end{equation}
The small constant $r_{0}$ is indicated in Fig.~\ref{fig3}. Let us chose the electric field at $t<0$ in the form
\begin{equation}
\label{16}
{\cal E}(t)=\frac{\hbar\Omega}{ar^{3}_{0}}F''\left[\frac{\Omega(t+t_{1})}{r^{2}_{0}}\right]
\end{equation}
where the function $F(z)=\exp(z)$ at $z\rightarrow -\infty$ and $F(z)=z$  at $z\rightarrow +\infty$. The moment $t_{1}$ is
\begin{equation}
\label{17}
t_{1}=\frac{ r_{0}}{\Omega}
\end{equation}
One can make the following transformation of the wave function in Eq.~(\ref{15}) \cite{PERELOMOV2}
\begin{equation}
\label{18}
\psi(x,t)=\varphi\left[x-\eta (t),t\right]\exp\left\{\frac{im}{\hbar}(x-\eta)\dot\eta+
\frac{1}{\hbar}\int^{t}dt_{1}\left[\frac{m}{2}{\dot\eta}^{2}+\eta\hspace{0.05cm}{\cal E}(t_{1})\right]\right\}
\end{equation}
where $\eta(t)$, which is an additional displacement of the wave packet under the action of ${\cal E}(t)$, satisfies the equation 
\begin{equation}
\label{19}
m\frac{\partial^{2}\eta}{\partial t^{2}}={\cal E}(t)
\end{equation}
The wave function $\varphi(y,t)$ is a solution of the Schr\"{o}dinger equation
\begin{equation}
\label{20}
i\hbar\hspace{0.1cm}\frac{\partial\varphi(y,t)}{\partial t}=-\frac{\hbar^{2}}{2m}\hspace{0.1cm}\frac{\partial^{2}\varphi}{\partial y^{2}}
+\frac{1}{r^{2}_{0}}V\left[\frac{y+\eta(t)}{r_{0}}\right]\varphi
\end{equation}
It can be seen from Eq.~(\ref{20}) that, as soon as the packet does not reach the barrier, the incident wave in the function $\varphi$ is 
violated weak and therefore the additional (due to the electric field) velocity of the incident wave is $\dot{\eta}$.

As follows from Eqs.~(\ref{19}) and (\ref{16}), the additional velocity of the packet is given by the relation 
\begin{equation}
\label{21}
m\frac{\partial\eta (t)}{\partial t}=\frac{\hbar}{ar_{0}}F'\left[\frac{\Omega(t+t_{1})}{r^{2}_{0}}\right]
\end{equation}
and the additional displacement is
\begin{equation}
\label{22}
\eta (t)=r_{0}a\hspace{0.1cm}\frac{\omega}{\Omega}F\left[\frac{\Omega(t+t_{1})}{r^{2}_{0}}\right]
\end{equation}
We consider $\Omega$ to be less than the intrinsic frequency $\omega$. The parameter $r_{0}$ is small. 

\begin{figure}[p]
\begin{center}
\vspace{1.5cm}
\leavevmode
\epsfxsize=\hsize
\epsfxsize=16cm
\epsfbox{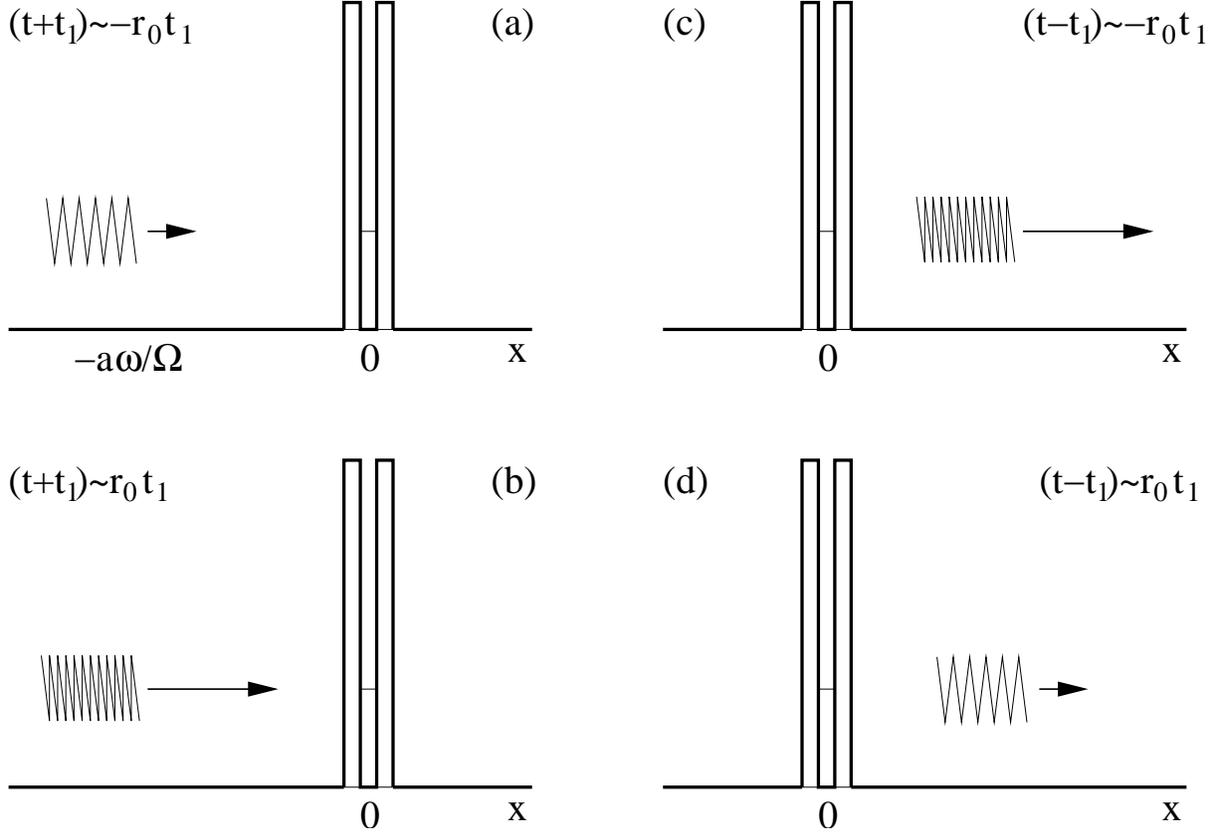}
\vspace{1cm}
\caption{A motion of the wave packet through the shrunk barrier. (a) the wave packet before acceleration (close to the moment 
$-t_{1}$) is at the distance $a\omega/\Omega$ from the barrier. (b) the wave packet after acceleration close to the moment $-t_{1}$. (c) 
the accelerated wave packet after penetration through the barrier close to the moment $t_{1}$. (d) the braked wave packet close to the 
moment $t_{1}$.}
\label{fig4}
\end{center}
\end{figure}

The wave packet, moving from the left, reaches the distance $a\omega/\Omega$ to the left from the barrier, at the moment $t=-t_{1}$ which
is shown in Fig.~\ref{fig4}(a). The electric field is switched on at the moment $-t_{1}$ and acts during the short time interval 
$r_{0}t_{1}$ accelerating the particle up to the velocity $\hbar/mar_{0}$ which follows from (\ref{21}). This is shown in 
Fig.~\ref{fig4}(b).
 
In the time interval between $-t_{1}$ and $t_{1}$ the motion of the packet occurs in the pure static potential $V(x/r_{0})/r^{2}_{0}$ with
the constant velocity $v=\hbar/mar_{0}$. The particle collides the barrier at the moment $t=0$ and than the process of quantum tunneling 
through the double barrier starts with filling up the resonance state $\psi_{R}$. This occurs if the energy 
$mv^{2}/2\sim\hbar^{2}/ma^{2}r^{2}_{0}$ coincides with some energy level $\sim\hbar^{2}/ma^{2}r^{2}_{0}$ in the well in Fig.~\ref{fig4}. 

Compared to Fig.~\ref{fig2}, the process is spatially compressed with the scale $r_{0}$. Fig.~\ref{fig2}(a) corresponds to 
Fig.~\ref{fig4}(b) and Fig.~\ref{fig2}(b) corresponds to Fig.~\ref{fig4}(c). Essentially, that the length of the exit packet in 
Fig.~\ref{fig4}(c) is not exponentially shorter then the enter packet in Fig.~\ref{fig4}(b). 

After the moment $t=t_{1}$ the reversed pulse of ${\cal E}(t)$ acts and brakes the particle down the initial velocity as in 
Fig.~\ref{fig4}(d). During the further motion, at the moment $t=t_{0}$, the scale $r(t)$ of the double barrier returns to its previous value
$r(t)=1$.

The dynamical electric field ${\cal E}(t)$ produces only exponentially weak processes of over-barrier photon assistance. To show this, let
us choose the variables $\xi =x/r_{0}$ and $\tau =t/r^{2}_{0}$. The renormalized electric field is 
$\varepsilon (\tau)={\cal E}(t)r^{3}_{0}$. In these terms the Schr\"{o}dinger equation (\ref{15}) takes the form
\begin{equation}
\label{23}
i\hbar\hspace{0.1cm}\frac{\partial\psi(\xi,\tau)}{\partial\tau}=-\frac{\hbar^{2}}{2m}\hspace{0.1cm}\frac{\partial^{2}\psi}{\partial\xi^{2}}
-\xi\varepsilon (\tau)\psi +V(\xi)\psi
\end{equation}
One can easily analyze the processes of quanta absorption due to the nonstationary field $\varepsilon (t)$ in Eq.~(\ref{23}) on the basis 
of Eq.~(\ref{14a}). The parameter $w\sim\sqrt{\omega/V_{0}}\ll 1$. Since $\hbar\Omega$ is much less than the barrier height $V_{0}$ the 
process of photon-assisted tunneling over the barrier is exponentially small in probability. 

As one can conclude, during a time interval, which is not exponentially long, the fraction (which is not exponentially small) of the 
incident wave packet penetrate the double barrier which always remains non-transparent according to WKB. The dynamical perturbations are 
always adiabatic in the sense that they vary slow (on the scale $r^{2}/\Omega$) compared to the intrinsic time of the system 
$r^{2}/\omega$.  
\section{DISCUSSION}
\label{sec:disc}
In this paper the dynamical potential barrier is constructed which satisfies the conditions (\ref{3a}) - (\ref{3c}), it is slow varying 
compared to the intrinsic classical frequency and the WKB transparency is always exponentially small. Nevertheless, the real transparency 
is not exponentially small which is an example of an extremely unusual phenomenon of quantum physics called Euclidean resonance. 

With the nonstationary potential proposed the dynamical problem is mapped on well known static resonant tunneling through a double barrier 
system with a resonant level inside. This happens on the time interval $\{-t_{1},t_{1}\}$ in Fig.~\ref{fig3}. When the energy of the 
incident wave coincides with the resonant level the transition becomes non-exponentially small. Using this example, one can follow a 
formation of the resonant state inside the barrier system. One can also observe that formation of the resonant state is a rigorously 
non-semiclassical process, related to a strong quantum coherence, as it was stated before \cite{IVLEV2,IVLEV4}. So, the dynamical 
potential, considered in the paper, reproduces the main features of Euclidean resonance.

We considered only the dynamical potential with two symmetric barriers as in Fig.~\ref{fig2}. But really one can start with a barrier of 
any shape at $t\rightarrow -\infty$ as in Fig.~\ref{1}(a) and slowly deform it to the double barrier form before the moment $-t_{0}$ in 
Fig.~\ref{fig3}. 

The dynamical potential proposed during its evolution becomes exponentially high and narrow and returns again to its initial shape. This 
may not have a direct connection to a real situation if the deformation is too strong. In this case it is not important since the goal is 
to construct a dynamical potential satisfying the conditions (\ref{3a}) - (\ref{3c}) in order to proof that the Schr\"{o}dinger equation 
allows ER solution in principal. 

For the particular nonstationary barrier, proposed in the paper, it follows that the phenomenon of Euclidean resonance is a dynamical 
analogue of static resonant tunneling. One can assume that for a general dynamical barrier ER also can be somehow mapped on static resonant
tunneling. 
\section{CONCLUSIONS}
An extremely unusual phenomenon of quantum physics (easy penetration of classical barriers) is studied. A certain nonstationary barrier is 
proposed with a very low WKB tunneling rate. The quantum dynamics of this barrier is mapped on resonant tunneling across a static double 
barrier with a resonant level inside. The real penetration through the dynamical barrier is not exponentially small providing an example of
Euclidean resonance. Other ER features are also manifested in that dynamical process, formation of the under-barrier resonant state and the
resonant character of the effect depending on the incident energy. Therefore, it is shown that the Schr\"{o}dinger equation allows 
solutions of the type of Euclidean resonance. It follows that Euclidean resonance is a dynamical analogue of static resonant tunneling.   
\acknowledgments
I am grateful to V.~Gudkov for valuable discussions.

\newpage

\end{document}